\documentclass[12pt,reqno]{amsart}
\allowdisplaybreaks[4]
\usepackage [latin1]{inputenc}
\usepackage{graphicx} %We can use any other package if it is necessary
\usepackage{amssymb}
\usepackage{color}
\usepackage{amsmath}
\usepackage{cite}
\usepackage{subfigure}
\usepackage{graphicx}
\usepackage{epstopdf}
\usepackage{bibentry}
\begin{document}

\begin{center}
{\large \sc \bf THE GENERALISED $\textbf{q}$-WRONSKIAN SOLUTIONS OF THE $\textbf{q}$-DEFORMED CONSTRAINED MODIFIED KP HIERARCHY}

\vskip 20pt

{\large Ge Yi,  Wei Wang,  Kelei Tian$^*$ and Ying Xu }

\vskip 20pt

{\it
School of Mathematics, Hefei University of Technology, Hefei 230601, China }

\bigskip

$^*$ Corresponding author:  {\tt kltian@ustc.edu.cn, kltian@hfut.edu.cn}

\bigskip

{\today}

\end{center}

\bigskip
\bigskip
\textbf{Abstract:} In this paper, we give the form of the $q$-cmKP hierarchy generated by the gauge transformation operator $T_{n+k}$. We show a necessary and sufficient condition to reduce the generalised $q$-Wronskian solutions from the $q$-mKP hierarchy to generalised the $q$-Wronskian solutions of $M$-component $q$-cmKP hierarchy.
\bigskip

\bigskip
\bigskip

\section{INTRODUCTION}
The quantum-calculus \cite{KlSc1,KaCh2,Exto3,Andr4} was used by some mathematicians to prove identities, such as the Ramanujan identity given in terms of $q$. In recent years, $q$-calculus has been used to construct quantum groups, quantum integrability and so on\cite{Sun5,Macf6,Ge7}. The $q$-deformed integrable system is an important branch of integrable system. Many quantum integrable systems have been relatively complete, such as the $q$-AKNS-D hierarchy, the $q$-KdV hierarchy \cite{Zhang8,Tu9,Frenkel10,Haine11,Khesin12}, the $q$-KP hierarchy \cite{Tu13,Iliev14,Iliev15,Tian16,Lin17}, sub-hierarchies of the $q$-KP hierarchy and its generalizing cases\cite{HeLi18,HeLi19,Geng20,Tian21,Cheng22,Cheng23,Mas24}. The wave functions, $\tau$ functions, additional symmetries and Virasoro constraints of these hierarchies have been studied.%\cite{Tu13,Iliev14,Iliev15}

The reduction of the $q$-Wronskian solutions of the $q$-mKP hierarchy to its constrained case has been studied. By using the eigenfunction symmetry constraints of the $q$-mKP hierarchy\cite{Yi25}, we show a necessary and sufficient condition to reduce the $q$-Wronskian solutions of the $q$-mKP hierarchy to the $q$-cmKP hierarchy. On the basis of the narrow $q$-Wronskian solutions of the $q$-cmKP hierarchy, it is extended to the generalized $q$-Wronskian solutions of the $q$-cmKP hierarchy, which plays a good supplementary role in exploring the physical significance of $q$-deformation. We can obtain the generalized $q$-Wonskian solutions of the $q$-mKP hierarchy via the gauge transformation of the $q$-mKP hierarchy \cite{Chau26,ChenGe27,ChenGe28,Cheng29,Oeve30,Cheng31}. Hence, we will continue to study the generalized $q$-Wonskian solutions of the $q$-cmKP hierarchy.

 The main purpose of this paper is to discuss how to reduce the $\tau$ function in generalised $q$-Wronskian forms of the $q$-mKP hierarchy to the solutions of the $M$-component $q$-cmKP hierarchy. We need to think about what the form of the $l$-constrained $q$-mKP hierarchy generated by the gauge transformation operator $T_{n+k}(n\geq k)$ is. We think about whether there is a necessary and sufficient condition for reducing the generalised $q$-Wronskian solutions of the $q$-mKP hierarchy to the $q$-cmKP hierarchy, and whether the results can be returned to the classical cmKP hierarchy as $q\to1$. In this paper, we have given the $l$-constrained mKP hierarchy generated by the gauge transformation operator $T_{n+k}(n\geq k)$. Finally, we provide an example to illustrate our results.

This paper is organized as follows. In Section 2, we introduce some basic results of $q$-mKP hierarchy. We give the $q$-cmKP hierarchy generated by the gauge transformation operator $T_{n+k}$. In Section 3, the necessary and sufficient condition reducing the $q$-Wronskian solutions of the $q$-mKP hierarchy to the solutions of the $q$-cmKP hierarchy is given which is the main result. In Section 4, we give a example with $T_{2+1}$. In Section 5, the conclusions are given.

\bigskip
\bigskip
\section{THE $q$-DEFORMED CONSTRAINED MODIFIED KP HIERARCHY GENERATED BY \textit{$T_{n+k}$}}
The $q$-derivative $\partial_{q}$ is defined by its actions on a function $f(x)$ as
\begin{eqnarray}
\partial_{q}(f(x))=\frac{f(qx)-f(x)}{x(q-1)}.
\end{eqnarray}
When $q\to1$, the $q$-derivative $\partial_{q}$ defined by the above equation is reduced to the derivative $\partial_{x}(f(x))$ in classical calculus theory.
The $q$-shift operator is defined as
\begin{eqnarray}
\theta(f(x))=f(qx).
\end{eqnarray}
 The $q$-shift operator $\theta$ and $\partial_{q}$ are not commutative and they satisfy
\begin{eqnarray}
q^{mk}\theta^{m}(\partial^{k}_{q}(f(x)))=\partial^{k}_{q}(\theta^{m}(f(x))),~~k,m\in \mathbb{Z}.\nonumber
\end{eqnarray}

Let $\partial^{-1}_{q}$ be the formal inverse operator of $\partial_{q}$, the algebraic multiplication of $\partial^{n}_{q}$ with the multiplication operator $f$ is given by the $q$-deformed Leibnitz rule
\begin{eqnarray}
\partial^{n}_{q}\circ f=\sum_{k\geq 0}\left(\begin{array}{c}n\\k
\end{array}\right)_{q}\theta^{n-k}(\partial^{k}_{q}(f))\partial^{n-k}_{q},~~n\in\mathbb{Z},
\end{eqnarray}
where the $q$-number and $q$-binomial are defined by
\begin{eqnarray}
\left(\begin{array}{c}n\\k
\end{array}\right)_{q}&=&\frac{(n)_{q}(n-1)_{q}\cdots(n-k+1)_{q}}{(1)_{q}(2)_{q}\cdots(k)_{q}},\nonumber\\
\left(\begin{array}{c}n\\0
\end{array}\right)_{q}&=&1,~~(n)_{q}=\frac{q^{n}-1}{q-1}.\nonumber
\end{eqnarray}

For a $q$-pseudo-differential operator  $A=\sum_{i}a_{i}\partial^{i}_{q}$, we divide operator $A$ into two parts. One part is $A_{\geq k}=\sum_{i\geq k}a_{i}\partial^{i}_{q}$, the other part is $A_{<k}=\sum_{i<k}a_{i}\partial^{i}_{q}$. The symbol $A(f)$ means the action of $A$ on $f$, whereas $Af$ or $A\circ f$ denotes the operator multiplication of $A$ and $f$. The conjugate operation $\ast$ has the rules $(AB)^{\ast}=B^{\ast}A^{\ast}$, $\partial^{\ast}_{q}=-\partial_{q}\theta^{-1}=-\frac{1}{q}\partial_{\frac{1}{q}}$ , $(\partial^{-1}_{q})^{\ast}=(\partial^{\ast}_{q})^{-1}=-\theta\partial^{-1}_{q}$ and $f^{*}=f$.

The $q$-exponent $e_{q}(x)$ is defined as
\begin{eqnarray}
e_{q}(x)=\sum\limits_{k=1}^{\infty}\frac{x^{n}}{(n)_{q}!}=exp(\sum\limits_{k=1}^{\infty}\frac{(1-q)^{k}}{k(1-q^{k})}x^{k}).\nonumber
\end{eqnarray}
The $e_{q}(x)$ satisfies
$\partial^{k}_{q}(e_{q}(xz))=z^{k}e_{q}(xz),\;\;\;k\in\mathbb{Z}.$
\\\\
\textbf{Lemma  1} For any $q$-pseudo-differential operator $A$ and arbitrary functions $f,g$ and $h$, here are some of operator identities used in this paper.
\begin{eqnarray}
%(f^{-1}Af)_{\geq 1}&=&f^{-1}A_{\geq 1}f-f^{-1}A_{\geq 1}(f),\\
%((\partial_{q}(f))^{-1}\partial_{q}\circ A\circ\partial^{-1}_{q}\partial_{q}(f))_{\geq 1}&=&(\partial_{q}(f))^{-1}\partial_{q}A_{\geq 1}\partial^{-1}_{q}\partial_{q}(f)%\nonumber\\
%-(\partial_{q}(f))^{-1}(\partial_{q}A_{\geq 1})(f),\\
%(\partial^{-1}_{q}fAf^{-1}\partial_{q})_{\geq 1}&=&\partial^{-1}_{q}fA_{\geq 1}f^{-1}\partial_{q}-\partial^{-1}_{q}f^{-1}A^{\ast}_{\geq 1}(f)\partial_{q},\\
(A\circ f\circ\partial^{-1}_{q})_{<0}&=&A_{\geq 0}(f)\circ\partial^{-1}_{q}+A_{<0}\circ f\circ\partial^{-1}_{q},\label{lem1}\\
(\partial^{-1}_{q}\circ f\circ A)_{<0}&=&\partial^{-1}_{q}\circ A^{\ast}_{\geq 0}(f)+\partial^{-1}_{q}\circ f\circ A_{<0},\label{lem2}\\
%f(x)\partial^{-1}_{q}&=&\sum\limits_{k=1}^{M}\frac{1}{q^{k}}\partial^{-k-1}_{q}\circ(\partial^{k}_{q}(f(qx))),\nonumber\\
f\partial^{-1}_{q}&=&\sum\limits_{k=1}^{M}\frac{1}{q^{k}}\partial^{-k-1}_{q}\circ(\partial^{k}_{q}(\theta (f))),\label{lem3}\\
\partial^{-1}_{q}\circ g\circ\partial^{-1}_{q}\circ h&=&\partial^{-1}_{q}(g)\circ\partial^{-1}_{q}\circ h-\partial^{-1}_{q}\circ\theta(\partial^{-1}_{q}(g))\circ h,\label{lem4}\\
(\partial^{-1}_{q}\circ f\circ\partial^{l}_{q}\circ g\circ\partial^{-1}_{q}\circ h)_{<0}
&=&\partial^{-1}_{q}(f\partial^{l}_{q}(g))\circ\partial^{-1}_{q}\circ h\nonumber\\
&&-(-1)^{l}q^{-(1+2+\cdots+l)}\partial^{-1}_{q}\circ\theta(\partial^{-1}_{q}(\theta^{-l}(\partial^{l}_{q}(f))g))\circ h.\label{lem5}
\end{eqnarray}\\
\textbf{Proof.} The formulas \eqref{lem1}-\eqref{lem4} which are derived from \cite{Cheng23} are operator identities. For the formula \eqref{lem5}, we have the following proof.
\begin{eqnarray}
(\partial^{-1}_{q}\circ f\circ\partial^{l}_{q}\circ g\circ\partial^{-1}_{q}\circ h)_{<0}&=&(\partial^{-1}_{q}\circ f\circ\partial^{l}_{q})_{<0}\circ g\circ\partial^{-1}_{q}\circ h+(\partial^{-1}_{q}\circ f\circ\partial^{l}_{q})_{\geq0}(g)\circ\partial^{-1}_{q}\circ h\nonumber,\\
%\partial^{-1}_{q}f\partial^{l}_{q}&=&\theta^{-1}(f)\partial^{l-1}_{q}-q^{-1}\theta^{-2}(\partial_{q}(f))\partial^{l-2}_{q}+\cdots+(-1)^{l-1}q^{-(1+2+\cdots+l-1)}\theta^{-l}(\partial^{l-1}_{q}(f))\nonumber\\
%&+&(-1)^{l}q^{-(1+2+\cdots+l-1+l)}\theta^{-l-1}(\partial^{l}_{q}(f))\partial_{q}+\cdots\nonumber\\
(\partial^{-1}_{q}\circ f\circ\partial^{l}_{q})_{<0}
&=&(-1)^{l}q^{-(1+2+\cdots+l)}\partial^{-1}_{q}\circ\theta^{-l}(\partial^{l}_{q}(f))\nonumber,\\
(\partial^{-1}_{q}\circ f\circ\partial^{l}_{q})_{\geq0}(g)
&=&\sum\limits_{i=0}^{l-1}(-1)^{i}q^{-(1+2+\cdots+i)}\theta^{-i-1}(\partial^{-i}_{q}(f))\partial^{l-i-1}_{q}(g)\nonumber,
\end{eqnarray}
so we have
\begin{eqnarray}
&&(\partial^{-1}_{q}\circ f\circ\partial^{l}_{q}\circ g\circ\partial^{-1}_{q}\circ h)_{<0}\nonumber\\
&=&(-1)^{l}q^{-(1+2+\cdots+l)}\partial^{-1}_{q}\circ\theta^{-l}(\partial^{l}_{q}(f))\circ g\circ\partial^{-1}_{q}\circ h\nonumber\\
&&+\sum\limits_{i=0}^{l-1}(-1)^{i}q^{-(1+2+\cdots+i)}\theta^{-i-1}(\partial^{-i}_{q}(f))\partial^{l-i-1}_{q}(g)\circ\partial^{-1}_{q}\circ h\nonumber\\
&=&(-1)^{l}q^{-(1+2+\cdots+l)}\partial^{-1}_{q}(\theta^{-l}(\partial^{l}_{q}(f))g)\circ\partial^{-1}_{q}-(-1)^{l}q^{-(1+2+\cdots+l)}\partial^{-1}_{q}\circ\theta(\partial^{-1}_{q}(\theta^{-l}(\partial^{l}_{q}(f))g))\circ h\nonumber\\
&&+\sum\limits_{i=0}^{l-1}(-1)^{i}q^{-(1+2+\cdots+i)}\theta^{-i-1}(\partial^{-i}_{q}(f))\partial^{l-i-1}_{q}(g)\circ\partial^{-1}_{q}\circ h\nonumber\\
&=&\partial^{-1}_{q}(f\partial^{l}_{q}(g))\circ\partial^{-1}_{q}\circ h-
(-1)^{l}q^{-(1+2+\cdots+l)}\partial^{-1}_{q}\circ\theta(\partial^{-1}_{q}(\theta^{-l}(\partial^{l}_{q}(f))g))\circ h.\nonumber
\end{eqnarray}
$\hfill\square $

The Lax equation of the $q$-mKP hierarchy is given by
\begin{eqnarray}
\frac{\partial L}{\partial t_{n}}=[B_{n},L],\;\;\;\;n=1,2,3,\ldots,\label{qmKPLax}
\end{eqnarray}
where $B_{n}=(L^{n})_{\geq 1}$. The Lax operator $L$ is defined by
\begin{eqnarray}
L=u_{1}\partial_{q}+u_{0}+u_{-1}\partial^{-1}_{q}+u_{-2}\partial^{-2}_{q}+\cdots,\label{Lax}
\end{eqnarray}
in which
$u_{i}=u_{i}(x,t)=u_{i}(x,t_{1},t_{2},\cdots)$.

The Lax operator $L$ of the $q$-mKP hierarchy can be expressed in terms of the dressing operator $Z$ as
\begin{eqnarray}
L=Z\circ\partial_{q}\circ Z^{-1}\nonumber
\end{eqnarray}
with the dressing operator $Z=z_{0}+z_{1}\partial^{-1}_{q}+z_{2}\partial^{-2}_{q}+\cdots$ ($z^{-1}_{0}$ exists).
The Lax equation \eqref{qmKPLax} is equivalent to the Sato equation
\begin{eqnarray}
\frac{\partial Z}{\partial t_{n}}=-(Z\circ\partial^{n}_{q}\circ Z^{-1})_{\leq 0}\circ Z.
\end{eqnarray}

Similar to the $q$-KP hierarchy, the definition of the $q$-wave function $W_{q}(x,t)$ and the $q$-adjoint wave function $W^\ast_{q}(x,t)$ can be given as
\begin{eqnarray}
W_{q}(x,t;\lambda)&=&Ze_{q}(x\lambda)exp(\sum\limits_{i=1}^{\infty}t_{i}\lambda^{i}),\nonumber\\
W^\ast_{q}(x,t;\lambda)&=&(Z^{-1}\partial^{-1}_{q})^\ast |_{\frac{x}{q}} e_{\frac{1}{q}}(-x\lambda)exp(-\sum\limits_{i=1}^{\infty}t_{i}\lambda^{i}),\nonumber
\end{eqnarray}
where
\begin{eqnarray}
P|_{\frac{x}{t}}=\sum_{i}p_{i}(\frac{x}{t})t^{i}\partial^{i}_{q}\nonumber
\end{eqnarray}
for a $q$-pseudo-differential operator $P=\sum_{i}p_{i}(x)\partial^{i}_{q}$. $W_{q}(x,t)$  satisfies the following equations
\begin{eqnarray}
LW_{q}&=&\lambda W_{q},\nonumber\\
\frac{\partial W_{q}}{\partial t_{n}}&=&B_{n}W_{q}.\nonumber
%(\partial_{q}L\partial^{-1}_{q}|_{\frac{x}{q}})^\ast W^\ast_{q}&=&\lambda W^\ast_{q},\;\;\;\;\;\;\;\frac{\partial W^\ast_{q}}{\partial t_{n}}=-((B_{n}|_{\frac{x}{q}})^\ast W^\ast_{q}).
\end{eqnarray}\\\\
\textbf{Lemma 2} \cite{Cheng32} Let $P$ and $Q$ be $q$-pseudo-differential operators, then one have
\begin{eqnarray}
res_{\lambda}(Pe_{q}(x\lambda)Q^\ast|_{\frac{x}{q}}e_{\frac{1}{q}}(-x\lambda))=res_{\partial_{q}}(PQ).
\end{eqnarray}
By using the above lemma, the bilinear identity of the $q$-mKP hierarchy  can be established.\\\\
\textbf{Lemma 3} \cite{Yi25} The bilinear identity
\begin{eqnarray}
res_{\lambda}\left((\partial^{n}_{q}\partial^{\alpha_{1}}_{t_{1}}\cdots\partial^{\alpha_{m}}_{t_{m}} W_{q})W^\ast_{q}\right)=1
\end{eqnarray}
holds for any $n\in\mathbb{Z}_{+}$ and $\alpha=(\alpha_{1},\cdots, \alpha_{m})\in \mathbb{Z}_{+}^{m}$.\\
%\textbf{Proof.} Since $\partial_{t_{i}}W_{q}=B_{i}(W_{q})$, we only need to prove the equation $res_{\lambda}(\partial^{n}_{q}\circ W_{q}W^\ast_{q})=1$,
%\begin{eqnarray}	
%res_{\lambda}(\partial^{n}_{q}\circ W_{q}W^\ast_{q})&=&res_{\lambda}(\partial^{n}_{q}\circ Ze_{q}(x\lambda)exp(\sum\limits_{i=1}^{\infty}t_{i}\lambda^{i})(Z^{-1}\partial^{-1}_{q}|_{\frac{x}{q}}^\ast e_{\frac{1}{q}}(-x\lambda)exp(-\sum\limits_{i=1}^{\infty}t_{i}\lambda^{i}))\nonumber\\
%&=&res_{\lambda}(\partial^{n}_{q}\circ Ze_{q}(x\lambda)(Z^{-1}\partial^{-1}_{q}|_{\frac{x}{q}})^\ast e_{\frac{1}{q}}(-x\lambda))\nonumber\\
%&=&res_{\partial_{q}}(\partial^{n}_{q}\circ Z Z^{-1}\circ\partial^{-1}_{q})\nonumber\\
%&=&res_{\partial_{q}}\partial^{n-1}_{q}\nonumber\\
%&=&1. \nonumber
%\end{eqnarray}

The eigenfunction $s$ and the adjoint eigenfunction $r$ of the $q$-mKP hierarchy satisfy the following equations
\begin{eqnarray}
\frac{\partial s}{\partial t_{n}}&=&(L^{n})_{\geq 1}(s),\\
\frac{\partial r}{\partial t_{n}}&=&-(L^{n})_{\geq 1}^{\ast}(r).
\end{eqnarray}
%Then we define the $k$-constraints of $q$-mKP hierarchy
%\begin{eqnarray}
%(L^{k})_{\leq 0}=\sum\limits_{i=1}^{M}\phi_{i}\circ\partial^{-1}_{q}\circ\psi_{i}\circ\partial_{q},
%\end{eqnarray}
%for suitable eigenfunction $\phi_{i}$ and adjoint eigenfunction $\psi_{i}$.\\

The solution of the Sato equation can be represented by a simple $\tau$ function, let $L^{l}=(L^{l})_{\le0}+(L^{l})_{>0}$, we know that the Lax operator \eqref{Lax} and generalized $l$-constraints
\begin{eqnarray}
(L^{l})_{\le0}=\sum\limits_{i=1}^{M}s_{i}\circ\partial^{-1}_{q}\circ r_{i}\circ\partial^{-1}_{q}
\end{eqnarray}
are compatible. And $L^{l}$ is the Lax operator of the $M$-component $q$-cmKP hierarchy.
\\

The gauge transformation operators \textit{\textbf{$T_{n+k}$}}\cite{Cheng23} can be given by the following lemma. \\\\
\textbf{Lemma 4} When $k<n$,
\begin{eqnarray}\label{Tnk}
T_{n+k}&=&\frac{1}{IW^{q}_{k,n+1}(\Psi_{k},\cdots,\Psi_{1};\Phi_{1},\cdots,\Phi_{n},1)} \nonumber \\
&&\times\begin {vmatrix}
\partial^{-1}_{q}(\Phi_{1}\Psi_{k})&\partial^{-1}_{q}(\Phi_{2}\Psi_{k})&\cdots&\partial^{-1}_{q}(\Phi_{n}\Psi_{k})&\partial^{-1}_{q}\circ\Psi_{k}\\
\partial^{-1}_{q}(\Phi_{1}\Psi_{k-1})&\partial^{-1}_{q}(\Phi_{2}\Psi_{k-1})&\cdots&\partial^{-1}_{q}(\Phi_{n}\Psi_{k-1})&\partial^{-1}_{q}\circ\Psi_{k-1}\\
\vdots&\vdots&\cdots&\vdots&\vdots\\
\partial^{-1}_{q}(\Phi_{1}\Psi_{1})&\partial^{-1}_{q}(\Phi_{2}\Psi_{1})&\cdots&\partial^{-1}_{q}(\Phi_{n}\Psi_{1})&\partial^{-1}_{q}\circ\Psi_{1}\\
\Phi_{1}&\Phi_{2}&\cdots&\Phi_{n}&1\\
\partial_{q}(\Phi_{1})&\partial_{q}(\Phi_{2})&\cdots&\partial_{q}(\Phi_{n})&\partial_{q}\\
\vdots&\vdots&\cdots&\vdots&\vdots\\
\partial^{n-k}_{q}(\Phi_{1})&\partial^{n-k}_{q}(\Phi_{2})&\cdots&\partial^{n-k}_{q}(\Phi_{n})&\partial^{n-k}_{q}\\
\end{vmatrix},
\end{eqnarray}
\begin{eqnarray}
T^{-1}_{n+k}&=&\frac{(-1)^{n-1}q^{-k}IW^{q}_{k,n+1}(\Psi_{k},\cdots,\Psi_{1};\Phi_{1},\cdots,\Phi_{n},1)}{ IW^{q}_{k,n}(\Psi_{k},\cdots,\Psi_{1};\Phi_{1},\cdots,\Phi_{n})\theta (IW^{q}_{k,n}(\Psi_{k},\cdots,\Psi_{1};\Phi_{1},\cdots,\Phi_{n}))}\nonumber\\
&\times&\begin {vmatrix}
\Phi_{1}\partial^{-1}_{q}&\theta(\partial^{-1}_{q}(\Psi_{k}\Phi_{1}))&\cdots&\theta(\partial^{-1}_{q}(\Psi_{1}\Phi_{1}))&\theta(\Phi_{1})&\cdots&\theta(\partial^{(n-k-2)}_{q}(\Phi_{1}))\\
\Phi_{2}\partial^{-1}_{q}&\theta(\partial^{-1}_{q}(\Psi_{k}\Phi_{2}))&\cdots&\theta(\partial^{-1}_{q}(\Psi_{1}\Phi_{2}))&\theta(\Phi_{2})&\cdots&\theta(\partial^{(n-k-2)}_{q}(\Phi_{2}))\\
\vdots&\vdots&\cdots&\vdots&\vdots&\cdots&\vdots\\
\Phi_{n}\partial^{-1}_{q}&\theta(\partial^{-1}_{q}(\Psi_{k}\Phi_{n}))&\cdots&\theta(\partial^{-1}_{q}(\Psi_{1}\Phi_{n}))&\theta(\Phi_{n})&\cdots&\theta(\partial^{(n-k-2)}_{q}(\Phi_{n}))
\end{vmatrix},\nonumber
\end{eqnarray}
and
\begin{eqnarray}
(T^{-1}_{n+k})^{*}&=&\frac{(-1)^{n}q^{-k}IW^{q}_{k,n+1}(\Psi_{k}, \cdots,\Psi_{1};\Phi_{1},\cdots,\Phi_{n},1)}{ IW^{q}_{k,n}(\Psi_{k},\cdots,\Psi_{1};\Phi_{1},\cdots,\Phi_{n})\theta (IW^{q}_{k,n}(\Psi_{k},\cdots,\Psi_{1};\Phi_{1},\cdots,\Phi_{n}))}\nonumber\\
&\times&\begin {vmatrix}
\theta\partial^{-1}_{q}\circ\Phi_{1}&\partial^{-1}_{q}\theta(\Psi_{k}\Phi_{1})&\cdots&\partial^{-1}_{q}\theta(\Psi_{1}\Phi_{1})&\theta(\Phi_{1})&\cdots&\theta(\partial^{(n-k-2)}_{q}(\Phi_{1}))\\
\theta\partial^{-1}_{q}\circ\Phi_{2}&\partial^{-1}_{q}\theta(\Psi_{k}\Phi_{2})&\cdots&\partial^{-1}_{q}\theta(\Psi_{1}\Phi_{2})&\theta(\Phi_{2})&\cdots&\theta(\partial^{(n-k-2)}_{q}(\Phi_{2}))\\
\vdots&\vdots&\cdots&\vdots&\vdots&\cdots&\vdots\\
\theta\partial^{-1}_{q}\circ\Phi_{n}&\partial^{-1}_{q}\theta(\Psi_{k}\Phi_{n})&\cdots&\partial^{-1}_{q}\theta(\Psi_{1}\Phi_{n})&\theta(\Phi_{n})&\cdots&\theta(\partial^{(n-k-2)}_{q}(\Phi_{n}))
\end{vmatrix}.\nonumber
\end{eqnarray}

 The determinant of \textit{$T_{n+k}$} in Lemma 4 is expanded by the last column and the functions are on the left-hand side. The determinant of \textit{$T^{-1}_{n+k}$} is expanded by the first column and the functions are on the right-hand side, and the coefficient function before the determinant should be placed after the operators $\Phi_{i}\partial^{-1}$. The determinant of \textit{$(T^{-1}_{n+k})^*$} is expanded by the first column and the functions are on the left-hand side.

The generalized Wronskian determinant is defined in the following form
\begin{eqnarray}
\tau^{q}_{k,n}&=&IW^{q}_{k,n}(\Psi_{k},\cdots,\Psi_{1};\Phi_{1},\cdots,\Phi_{n})\nonumber\\
&=&\begin {vmatrix}
\partial^{-1}_{q}(\Phi_{1}\Psi_{k})&\partial^{-1}_{q}(\Phi_{2}\Psi_{k})&\cdots&\partial^{-1}_{q}(\Phi_{n}\Psi_{k})\\
\vdots&\vdots&\cdots&\vdots\\
\partial^{-1}_{q}(\Phi_{1}\Psi_{1})&\partial^{-1}_{q}(\Phi_{2}\Psi_{1})&\cdots&\partial^{-1}_{q}(\Phi_{n}\Psi_{1})\\
\Phi_{1}&\Phi_{2}&\cdots&\Phi_{n}\\
\partial_{q}(\Phi_{1})&\partial_{q}(\Phi_{2})&\cdots&\partial_{q}(\Phi_{n})\\
\vdots&\vdots&\cdots&\vdots\\
\partial^{n-k-1}_{q}(\Phi_{1})&\partial^{n-k-1}_{q}(\Phi_{2})&\cdots&\partial^{n-k-1}_{q}(\Phi_{n})
\end{vmatrix}.\nonumber
\end{eqnarray}
\textbf{Remark 1} When $k=0$,\\
\begin{eqnarray}
IW^{q}_{0,n}=W^{q}_{n}(\Phi_{1},\cdots,\Phi_{n})
=\begin {vmatrix}
\Phi_{1}&\Phi_{2}&\cdots&\Phi_{n}\\
\partial_{q}(\Phi_{1})&\partial_{q}(\Phi_{2})&\cdots&\partial_{q}(\Phi_{n})\\
\vdots&\vdots&\cdots&\vdots\\
\partial^{n-1}_{q}(\Phi_{1})&\partial^{n-1}_{q}(\Phi_{2})&\cdots&\partial^{n-1}_{q}(\Phi_{n})
\end{vmatrix}.\nonumber
\end{eqnarray}
Thus \textit{$T_{n+k}$} and \textit{$T^{-1}_{n+k}$} have the following forms
\begin{eqnarray}
&&T_{n+k}=\sum_{p=0}^{n-k}a_{p}\partial_{q}^{p}+\sum_{p=-1}^{-k}
a_{p}\partial^{-1}_{q}\circ\Psi_{|p|}, \\
&&T^{-1}_{n+k}=\sum_{j=1}^{n}\Phi_{j}\partial^{-1}_{q}\circ b_{j}.
\end{eqnarray}

\textbf{Remark 2} When $k=n$,
\begin{eqnarray}
T_{n+n}=&&\frac{1}{IW^{q}_{n,n}(\Psi_{n},\cdots,\Psi_{1};\Phi_{1},\cdots,\Phi_{n})}\nonumber\\
&&\times\begin {vmatrix}
\partial^{-1}_{q}(\Phi_{1}\Psi_{n})&\partial^{-1}_{q}(\Phi_{2}\Psi_{n})&\cdots&\partial^{-1}_{q}(\Phi_{n}\Psi_{n})&\partial^{-1}_{q}\circ\Psi_{n}\\
\vdots&\vdots&\cdots&\vdots\\
\partial^{-1}_{q}(\Phi_{1}\Psi_{1})&\partial^{-1}_{q}(\Phi_{2}\Psi_{1})&\cdots&\partial^{-1}_{q}(\Phi_{n}\Psi_{1})&\partial^{-1}_{q}\circ\Psi_{1}\\
\Phi_{1}&\Phi_{2}&\cdots&\Phi_{n}&1
\end{vmatrix},\nonumber
\end{eqnarray}
and
\begin{eqnarray}
T_{n+n}=1+\sum_{p=-1}^{-k}
a_{p}\partial^{-1}_{q}\circ\Psi_{|p|}.\nonumber
\end{eqnarray}

In other words, $T_{n+k}$ can still be expressed by the formula \eqref{Tnk}.\\\\
\textbf{Remark 3} There is a sufficient condition $IW^{q}_{k,n}\neq0$ for the existence of these operators, in addition, the eigenfunctions $\Phi_{i}, (i=1,2,\cdots,n)$ and the adjoint eigenfunctions $\Psi_{j}, (j=1,2,\cdots,k)$ of the $q$-mKP hierarchy defined by the free operators $L^{0}=\partial_{q}$ satisfy
\begin{eqnarray}
&&\frac{\partial\Phi_{i}}{\partial t_{n}}=B^{0}_{n}(\Phi_{i})=\partial^{n}_{q}(\Phi_{i}),\\
&&\frac{\partial\Psi_{j}}{\partial t_{n}}=-(\partial_{q}B^{0}_{n}\partial^{-1}_{q})^{*}(\Psi_{j}).
\end{eqnarray}

Let's first prove that \textit{$T_{n+k}$} is dressing operator of the $q$-mKP hierarchy.\\\\
\textbf{Proposition 1} \textit{$T_{n+k}$} satisfies Sato equation
\begin{eqnarray}
(T_{n+k})_{t_{n}}%=\frac{\partial T_{n+k}}{\partial t_{n}}
=-(T_{n+k}\circ\partial^{n}_{q}\circ T_{n+k}^{-1})_{\leq0}T_{n+k}.
\end{eqnarray}\\
\textbf{Proof.} Let $l=1$, and let $L=T_{n+k}\circ\partial_{q}\circ T_{n+k}^{-1}=u_{1}\partial_{q}+u_{0}+u_{-1}\partial^{-1}_{q}+u_{-2}\partial^{-2}_{q}+u_{-3}\partial^{-3}_{q}+\cdots$, $L$ satisfies
\begin{eqnarray}
\frac{\partial L}{\partial t_{n}}=[B_{n},L]=[L,(L^{n})_{\leq0}].\nonumber
\end{eqnarray}
Thus
\begin{eqnarray}
\frac{\partial L}{\partial t_{n}}&=&\frac{\partial }{\partial t_{n}}(T_{n+k}\circ\partial_{q}\circ T_{n+k}^{-1})\nonumber\\
%&=&\frac{\partial T_{n+k}}{\partial t_{n}}\circ\partial_{q}\circ T_{n+k}^{-1}+T_{n+k}\circ\partial_{q}\circ\frac{\partial T^{-1}_{n+k}}{\partial t_{n}}\nonumber\\
&=&(T_{n+k})_{t_{n}}\circ\partial_{q}\circ T_{n+k}^{-1}+T_{n+k}\circ\partial_{q}\circ(T^{-1}_{n+k})_{t_{n}}\nonumber\\
&=&(T_{n+k})_{t_{n}}\circ\partial_{q}\circ T_{n+k}^{-1}-T_{n+k}\circ\partial_{q}\circ T^{-1}_{n+k}\circ(T_{n+k})_{t_{n}}\circ T^{-1}_{n+k}\nonumber\\
&=&(T_{n+k})_{t_{n}}\circ T^{-1}_{n+k}\circ T_{n+k} \circ\partial_{q}\circ T_{n+k}^{-1}-T_{n+k} \circ\partial_{q}\circ   T_{n+k}^{-1}\circ(T_{n+k})_{t_{n}}\circ T^{-1}_{n+k}\nonumber\\
&=&-[L,(T_{n+k})_{t_{n}}T^{-1}_{n+k}],\nonumber
\end{eqnarray}
and we have
\begin{eqnarray}
%&&(L^{n})_{\le0}=-(T_{n+k})_{t_{n}}T^{-1}_{n+k}\\
&&(T_{n+k})_{t_{n}}=-(T_{n+k}\circ\partial^{n}_{q}\circ T_{n+k}^{-1})_{\leq0}T_{n+k}.\nonumber
\end{eqnarray}
$\hfill\square $

According to \cite{HeLi18}, the $\tau$ function of $q$-mKP hierarchy generated by $L=T_{n+k}\circ\partial_{q}\circ T_{n+k}^{-1}$ is generalized Wronskian $IW^{q}_{k,n}$. In addition, since $L^{l}=T_{n+k}\circ\partial^{l}_{q}\circ T_{n+k}^{-1}$ and the determinant of the gauge transformations, there is the following theorem.\\\\
\textbf{Theorem 1} The $q$-cmKP hierarchy generated by the gauge transformation \textit{$T_{n+k}$} is as follow
\begin{eqnarray}\label{theo1}
%L^{l}&=&(L^{l})_{>0}+(L^{l})_{\leq 0}\nonumber\\
(L^{l})_{\leq 0}&=&%(L^{l})_{\geq 1}
-\sum_{j=1}^{n}T_{n+k}(\partial^{l}_{q}(\Phi_{j}))\circ\partial^{-1}_{q}\circ \theta(\partial^{-1}_{q}(b_{j}))\circ\partial_{q}\nonumber\\
&&+(-1)^{l}q^{-(1+2+\cdots+l)}\sum_{p=-1}^{-k}a_{p}\circ\partial^{-1}_{q}\circ(T_{n+k}^{-1}\partial^{-1}_{q})^{*}(\theta^{-l}
(\partial^{l}_{q}(\Psi_{|p|})))\partial_{q}\nonumber\\
&&+\sum_{p=-1}^{-k}\sum_{j=1}^{n}a_{p}\circ\partial^{-1}_{q}(\Psi_{|p|}\partial^{l}_{q}(\Phi_{j}\partial^{-1}_{q}
(b_{j}))).
\end{eqnarray}
\textbf{Proof.} Using the identities in Lemma 1, we have
\begin{eqnarray}
&&(T_{n+k}\circ\partial^{l}_{q}\circ T_{n+k}^{-1})_{\leq 0}\nonumber\\
&=&((T_{n+k})_{+}\circ\partial^{l}_{q}\circ T_{n+k}^{-1}+(T_{n+k})_{-}\circ\partial^{l}_{q}\circ T_{n+k}^{-1})_{\leq 0}\nonumber\\
&=&((T_{n+k})_{+}\circ\partial^{l}_{q}\circ\sum_{j=1}^{n}\Phi_{j}\circ\partial^{-1}_{q}\circ b_{j}\circ\partial^{-1}_{q}+\sum_{p=-1}^{-k}a_{p}
\partial^{-1}_{q}\circ\Psi_{|p|}\circ\partial^{l}_{q}\circ\sum_{j=1}^{n}\Phi_{j}\circ\partial^{-1}_{q}\circ b_{j}\circ\partial^{-1}_{q})_{<0}\partial_{q}\nonumber\\
&=&-\sum_{j=1}^{n}(T_{n+k})_{+}(\partial^{l}_{q}(\Phi_{j}))\circ\partial^{-1}_{q}\circ\theta(\partial^{-1}_{q}(b_{j}))\partial_{q}\nonumber\\ &&+\sum_{p=-1}^{-k}\sum_{j=1}^{n}(a_{p}\circ\partial^{-1}_{q}\circ\Psi_{|p|}\circ\partial^{l}_{q}\circ\Phi_{j}\circ\partial^{-1}_{q}(b_{j})\partial^{-1}_{q})
_{<0}\partial_{q}\nonumber\\
&&-\sum_{p=-1}^{-k}\sum_{j=1}^{n}(a_{p}\circ\partial^{-1}_{q}\circ\Psi_{|p|}\circ\partial^{l}_{q}\circ\Phi_{j}\circ\partial^{-1}_{q}\circ\theta(\partial^{-1}_{q}(b_{j})))_{<0}\partial_{q}\nonumber\\
%&=&-\sum_{j=1}^{n}(T_{n+k})_{+}(\partial^{l}_{q}(\Phi_{j}))\circ\partial^{-1}_{q}\circ\theta(\partial^{-1}_{q}(b_{j}))\partial_{q}\nonumber\\
%&&+\sum_{p=-1}^{-k}\sum_{j=1}^{n}(a_{p}\circ\partial^{-1}_{q}(\Psi_{|p|}\circ\partial^{l}_{q}(\Phi_{j}\partial^{-1}_{q}(b_{j})))\partial^{-1}_{q})\partial_{q}\nonumber\\
%&&-(-1)^{l}q^{-(1+2+\cdots+l)}\sum_{p=-1}^{-k}\sum_{j=1}^{n}(a_{p}\circ\partial^{-1}_{q}\circ\theta(\partial^{-1}_{q}(\theta^{-l}
%(\partial^{l}_{q}(\Psi_{|p|}))\Phi_{j}\partial^{-1}_{q}(b_{j}))))\partial_{q}\nonumber\\
%&&-\sum_{p=-1}^{-k}\sum_{j=1}^{n}(a_{p}\circ\partial^{-1}_{q}(\Psi_{|p|}\circ\partial^{l}_{q}(\Phi_{j}))\circ\partial^{-1}_{q}\circ\theta(\partial^{-1}_{q}(b_{j})))\partial_{q}\nonumber\\
%&&+(-1)^{l}q^{-(1+2+\cdots+l)}\sum_{j=1}^{n}\sum_{p=-1}^{-k}(a_{p}\circ\partial^{-1}_{q}\circ\theta(\partial^{-1}_{q}(\theta^{-l}
%(\partial^{l}_{q}(\Psi_{|p|}))\Phi_{j}))\circ\theta(\partial^{-1}_{q}(b_{j})))\partial_{q}\nonumber\\
&=&-\sum_{j=1}^{n}T_{n+k}(\partial^{l}_{q}(\Phi_{j}))\circ\partial^{-1}_{q}\circ \theta(\partial^{-1}_{q}(b_{j}))\circ\partial_{q}\nonumber\\
&&+(-1)^{l}q^{-(1+2+\cdots+l)}\sum_{p=-1}^{-k}a_{p}\circ\partial^{-1}_{q}\circ(T_{n+k}^{-1}\partial^{-1}_{q})^{*}(\theta^{-l}
(\partial^{l}_{q}(\Psi_{|p|})))\partial_{q}\nonumber\\
&&+\sum_{p=-1}^{-k}\sum_{j=1}^{n}a_{p}\circ\partial^{-1}_{q}(\Psi_{|p|}\partial^{l}_{q}(\Phi_{j}\partial^{-1}_{q}
(b_{j}))).\nonumber
\end{eqnarray}
It is similar for the case of $k=n$.
$\hfill\square $

\section{REDUCED TO THE $M$-COMPONENT $q$-CMKP HIERARCHY}
According to Theorem 1, from the equation \eqref{theo1}, the $M$-component $q$-cmKP hierarchy can be given, then $(L^{l})_{\le0}$ can be expressed as
\begin{eqnarray}\label{L24}
(L^{l})_{\le0}&=&\sum_{i=1}^{M}s_{i}\circ\partial^{-1}_{q}\circ r_{i}\circ\partial_{q}\nonumber\\
&=&\sum_{i=1}^{\alpha}s_{i}\circ\partial^{-1}_{q}\circ r_{i}\circ\partial_{q}+\sum_{i=\alpha+1}^{\alpha+\beta=M-1}s_{i}\circ\partial^{-1}_{q}\circ r_{i}\circ\partial_{q}+s_{M}\circ\partial^{-1}_{q}\circ r_{M}\circ\partial_{q}\nonumber\\
&=&(L_{\alpha})_{\le0}+(L_{\beta})_{\le0}+(L_{\gamma})_{\le0},
\end{eqnarray}
here $(L_{\gamma})_{\le0}=s_{M}\circ\partial^{-1}_{q}\circ r_{M}\circ\partial_{q}=\sum_{p=-1}^{-k}\sum_{j=1}^{n}a_{p}\partial^{-1}_{q}(\Psi_{|p|}\partial^{l}_{q}(\Phi_{j}\partial^{-1}_{q}
(b_{j})))$, thus $r_{M}=1, s_{M}=\sum_{p=-1}^{-k}\sum_{j=1}^{n}a_{p}\partial^{-1}_{q}(\Psi_{|p|}\partial^{l}_{q}(\Phi_{j}\partial^{-1}_{q}
(b_{j})))$. $(L_{\alpha})_{\le0}$ and $(L_{\beta})_{\le0}$ are obtained by reducing the first and second parts of the right-hand side of the equation \eqref{theo1}, respectively.\\

Now, let's reduce $(L_{\alpha})_{\le0}$ and $(L_{\beta})_{\le0}$ respectively.\\\\
\textbf{Proposition 2} For the above Lax operator $L$, the following two conclusions are true.\\
\textbf{(I).} For $(L_{\alpha})_{\le0}$, if the Wronskian determinant $W_{\alpha}=(1,s_{1},\cdots,s_{\alpha})\neq0$, there is a unique $\alpha$-order $q$-differential operator
\begin{eqnarray}
A=\partial^{\alpha}_{q}+a_{\alpha-1}\partial^{\alpha-1}_{q}+\cdots+a_{1}\partial_{q}+1,
\end{eqnarray}
such that $(A(L_{\alpha})_{\le0})_{\le0}=0$. \\
\textbf{(II).} For $(L_{\beta})_{\le0}$, If the Wronskian determinant $W_{\alpha+1} =(1,r_{\alpha+1},\cdots,r_{\alpha+\beta})\neq0$, there exists
 $\beta$-order $q$-differential operator
 \begin{eqnarray}
 B=\partial^{\beta}_{q}+d_{\beta-1}\partial^{\beta-1}_{q}+\cdots+d_{1}\partial_{q}+1,
 \end{eqnarray}
 such that $((L_{\beta})_{\le0}B))_{\le0}=0$.\\\\
\textbf{Proof.} \textbf{(I)} By using Lemma 1,
\begin{eqnarray}
(A(L_{\alpha})_{\le0})_{\le0}&=&(A\sum_{i=1}^{\alpha}s_{i}\circ\partial^{-1}_{q}\circ r_{i}\circ\partial_{q})_{\le0}\nonumber\\
&=&\sum_{i=1}^{\alpha}A(s_{i})\circ\partial^{-1}_{q}\circ r_{i}\circ\partial_{q}=0.\nonumber
\end{eqnarray}
thus
\begin{eqnarray}
\begin{cases}a_{1}\partial_{q}(s_{1})+a_{2}\partial^{2}_{q}(s_{1})+\cdots+a_{\alpha}\partial^{\alpha}_{q}(s_{1})=-s_{1},\nonumber\\a_{1}\partial_{q}(s_{2})+a_{2}\partial^{2}_{q}(s_{2})+\cdots+a_{\alpha}\partial^{\alpha}_{q}(s_{2})=-s_{2},\nonumber\\\cdots\\a_{1}\partial_{q}(s_{\alpha})+a_{2}\partial^{2}_{q}(s_{\alpha})+\cdots+a_{\alpha}\partial^{\alpha}_{q}(s_{\alpha})=-s_{\alpha}.\nonumber
\end{cases}
\end{eqnarray}
By solving this liner equations, we know that when $W^{q}_{\alpha+1}(1,s_{1},s_{2}\cdots,s_{\alpha})\neq0$, $A$ is unique. There is the similar way for the conclusion  \textbf{(II)}.
$\hfill\square $
\\
\\
\textbf{Theorem 2} The $q$-mKP hierarchy of $L=T_{n+k}\partial_{q}T^{-1}_{n+k}$, with solutions given by $\tau_{0}$-functions $\tau_{0}=W_{\alpha+1} (1,s_{1},\cdots,s_{\alpha})\neq0$, if and only if
 \begin{eqnarray}\label{theo2}
 W^{q}_{\alpha+1}(T_{n+k}(\partial^{l}_{q}(\Phi_{j_{1}})),T_{n+k}(\partial^{l}_{q}(\Phi_{j_{2}})),\cdots,T_{n+k}(\partial^{l}_{q}(\Phi_{j_{\alpha+1}})))=0,
 \end{eqnarray}
 we have $(L_{\alpha})_{\le0}=\sum_{i=1}^{\alpha}s_{i}\circ\partial^{-1}_{q}\circ r_{i}\circ\partial_{q}$.\\\\
\textbf{Proof.} When $\alpha\leq n$, if $(L_{\alpha})_{\le0}=-\sum_{j=1}^{n}T_{n+k}(\partial^{l}_{q}(\Phi_{j}))\circ\partial^{-1}_{q}\circ \theta(\partial^{-1}_{q}(b_{j}))\partial_{q}=\sum_{i=1}^{\alpha}s_{i}\circ\partial^{-1}_{q}\circ r_{i}\circ\partial_{q}$ holds,  there exists an $\alpha$-order differential operator $A$ such that
\begin{eqnarray}
0=(A\circ(L_{\alpha})_{\le0})_{\le0}=-\sum_{j=1}^{n}A(T_{n+k}(\partial^{l}_{q}(\Phi_{j})))\circ\partial^{-1}_{q}\circ \theta(\partial^{-1}_{q}(b_{j}))\circ\partial_{q},\nonumber
\end{eqnarray}
so
\begin{eqnarray}
A(T_{n+k}(\partial^{l}_{q}(\Phi_{j})))=0,\quad j=1,2,\cdots,n,\nonumber
\end{eqnarray}
we know that $T_{n+k}(\partial^{l}_{q}(\Phi_{j})) \in ker(A)$, Since the kernel of $A$ is $\alpha$ dimensional space, at most $\alpha$ of these functions $T_{n+k}(\partial^{l}_{q}(\Phi_{j}))$ are linearly independent. Thus
\begin{eqnarray}
 W^{q}_{\alpha+1}(T_{n+k}(\partial^{l}_{q}(\Phi_{j_{1}})),T_{n+k}(\partial^{l}_{q}(\Phi_{j_{2}})),\cdots,T_{n+k}(\partial^{l}_{q}(\Phi_{j_{\alpha+1}})))=0.\nonumber
\end{eqnarray}
Conversely, when equation \eqref{theo2} holds, there are at most $\alpha$ functions in $T_{n+k}(\partial^{l}_{q}(\Phi_{j}))$ that are linearly independent. Then we can find suitable $\alpha$ functions $s_{1},s_{2},\cdots,s_{\alpha}$, such that
\begin{eqnarray}\label{sss}
T_{n+k}(\partial^{l}_{q}(\Phi_{j}))=-\sum_{i=1}^{\alpha}c_{ji}s_{i},\quad j=1,2,\cdots,n.
\end{eqnarray}
thus
\begin{eqnarray}\label{lsss}
(L_{\alpha})_{\le0}&=&-\sum_{j=1}^{n}T_{n+k}(\partial^{l}_{q}(\Phi_{j}))\circ\partial^{-1}_{q}\circ \theta(\partial^{-1}_{q}(b_{j}))\circ\partial_{q}\nonumber\\
&=&\sum_{j=1}^{n}\sum_{i=1}^{\alpha}c_{ji}s_{i}\circ\partial^{-1}_{q}\circ\theta(\partial^{-1}_{q}(b_{j}))\circ\partial_{q}\nonumber\\
&=&\sum_{i=1}^{\alpha}s_{i}\circ\partial^{-1}_{q}\circ(\sum_{j=1}^{n}c_{ji}\theta(\partial^{-1}_{q}(b_{j})))\circ\partial_{q}\nonumber\\
&=&\sum_{i=1}^{\alpha}s_{i}\circ\partial^{-1}_{q}\circ r_{i}\circ\partial_{q}.\nonumber
\end{eqnarray}
In this way we achieve the goal of reduction, and $r_{i}=\sum_{j=1}^{n}c_{ji}\theta(\partial^{-1}_{q}(b_{j}))$.\\
Using $q-$Wronskian determinant [6], we have
\begin{eqnarray}
IW_{k,n+1+\alpha+1}=(\Psi_{k},\Psi_{k-1},\cdots,\Psi_{1};\Phi_{1},\cdots,\Phi_{n},1,\partial^{l}_{q}(\Phi_{j_{1}}),\cdots,\partial^{l}_{q}(\Phi_{j_{\alpha+1}}))=0.
\end{eqnarray}
$\hfill\square $
\\
 \textbf{Theorem 3}  For $(L_{\beta})_{\le0}$, there is a $\beta$-order differential operator $B=\partial^{\beta}_{q}+d_{\beta-1}\partial^{\beta-1}_{q}+\cdots+d_{1}\partial_{q}+1$, and $\tau_{1}=W^{q}_{\beta+1}(1,r_{\alpha+1},\cdots,r_{\alpha+\beta})\neq0$, if and only if
 \begin{eqnarray}\label{theo3}
 W^{q}_{\beta+1}(\tilde{r}_{i_{1}},\cdots,\tilde{r}_{i_{\beta+1}})=0,\quad n+1\le i_{1},\cdots,i_{\beta+1}\le n+k,
 \end{eqnarray}
 where $\tilde{r}_{ij}=(T_{n+1}^{-1}\partial_{q}^{-1})^{*}\left( \theta^{-1} \left(\partial_{q}^{-1}(\Psi_{ij}) \right)\right) $, we have $(L_{\beta})_{\le0}=\sum_{i=\alpha+1}^{\alpha+\beta=M-1}s_{i}\circ\partial^{-1}_{q}\circ r_{i}\circ\partial_{q}$.\\\\
 \textbf{Proof.} When $\beta\leq k$, if
  \begin{eqnarray}
 (L_{\beta})_{\le0}&=&(-1)^{l}q^{-(1+2+\cdots+l)}\sum_{p=-1}^{-k}a_{p}\partial^{-1}_{q}\circ(T_{n+k}^{-1}\partial^{-1}_{q})^{*}(\theta^{-l}
 (\partial^{l}_{q}(\Psi_{|p|})))\circ\partial_{q}\nonumber\\
 &=&\sum_{i=\alpha+1}^{\alpha+\beta=M-1}s_{i}\circ\partial^{-1}_{q}\circ r_{i}\circ\partial_{q}\nonumber
  \end{eqnarray}
  holds, we can get by Proposition 2
 \begin{eqnarray}
 0&=&((L_{\beta})_{\le0}B))_{\le0}\nonumber\\
 &=&(-1)^{l}q^{-(1+2+\cdots+l)}[\sum_{p=-1}^{-k}a_{p}\partial^{-1}_{q}\circ(T_{n+k}^{-1}\partial^{-1}_{q})^{*}(\theta^{-l}
 (\partial^{l}_{q}(\Psi_{|p|})))\circ\partial_{q}\circ B]_{\le0}\nonumber\\
 &=&(-1)^{l}q^{-(1+2+\cdots+l)}[\sum_{p=-1}^{-k}a_{p}\partial^{-1}_{q}\circ\tilde{B}^{*}(T_{n+k}^{-1}\partial^{-1}_{q})^{*}(\theta^{-l}
 (\partial^{l}_{q}(\Psi_{|p|})))]\circ\partial_{q}.\nonumber
 \end{eqnarray}
From the above equation, we have
\begin{eqnarray}
\tilde{B}^{*}((T_{n+k}^{-1}\partial^{-1}_{q})^{*}(\theta^{-l}
(\partial^{l}_{q}(\Psi_{|
p|}))))=0.\nonumber
\end{eqnarray}
let $\tilde{r}_{n+|p|}=(T_{n+k}^{-1}\partial^{-1}_{q})^{*}(\theta^{-l}
(\partial^{l}_{q}(\Psi_{|p|})))$, we get that
\begin{eqnarray}
\tilde{B}^{*}(\tilde{r}_{n+1})=0,\quad
\tilde{B}^{*}(\tilde{r}_{n+2})=0,\quad\cdots,\quad \tilde{B}^{*}(\tilde{r}_{n+k})=0,\nonumber
\end{eqnarray}
so $\tilde{r}_{n+|p|}$ is the kernel of $\tilde{B}^{*}$. The the kernel of $\tilde{B}^{*}$ is $\beta$ dimensions, at most $\beta$ functions in $\tilde{r}_{n+|p|}$ are linearly independent. We know $W^{q}_{\beta+1}(\tilde{r}_{i_{1}},\cdots,\tilde{r}_{i_{\beta+1}})=0,\quad n+1\le i_{1},\cdots,i_{\beta+1}\le n+k$ is true.

When equation \eqref{theo3} holds, function $\tilde{r}_{n+|p|}$ can be expressed by the $\beta$ functions $r_{\alpha+1},r_{\alpha+2},\cdots r_{\alpha+\beta}$,
\begin{eqnarray}
\tilde{r}_{n+|p|}=\sum_{i=\alpha+1}^
{\alpha+\beta=M-1}v_{n+|p|,i}r_{i},\quad p=-1,-2,\cdots,-k,
\end{eqnarray}
thus
\begin{eqnarray}
(L_{\beta})_{\le0}&=&(-1)^{l}q^{-(1+2+\cdots+l)}\sum_{p=-1}^{-k}a_{p}\circ\partial^{-1}_{q}\circ(T_{n+k}^{-1}\partial^{-1}_{q})^{*}(\theta^{-l}
(\partial^{l}_{q}(\Psi_{|p|})))\circ\partial_{q}\nonumber\\
&=&\sum_{i=\alpha+1}^
{\alpha+\beta=M-1}(-1)^{l}q^{-(1+2+\cdots+l)}\sum_{p=-1}^{-k}a_{p}v_{n+|p|,i}\circ\partial^{-1}_{q}\circ r_{i}\circ\partial_{q}\nonumber\\
&=&\sum_{i=\alpha+1}^{\alpha+\beta=M-1}s_{i}\circ\partial^{-1}_{q}\circ r_{i}\circ\partial_{q},\nonumber
\end{eqnarray}
here $s_{i}=(-1)^{l}q^{-(1+2+\cdots+l)}\sum_{p=-1}^{-k}a_{p}v_{n+|p|,i}$.
$\hfill\square $
\\
\section{EXAMPLE}

Let's take the $q$-cmKP hierarchy generated by $T_{2+1}$ as an example.
When
\begin{eqnarray}
IW^{q}_{1,2}(\Psi_{1};\Phi_{1},\Phi_{2})=
\begin{vmatrix}
\partial^{-1}_{q}(\Phi_{1}\Psi_{1})& \partial^{-1}_{q}(\Phi_{2}\Psi_{1})\\
\Phi_{1}&\Phi_{2}
\end{vmatrix}\neq0,\nonumber
\end{eqnarray}
\begin{eqnarray}
IW^{q}_{1,2+1}(\Psi_{1};\Phi_{1},\Phi_{2},1)=
\begin{vmatrix}
\partial^{-1}_{q}(\Phi_{1}\Psi_{1})& \partial^{-1}_{q}(\Phi_{2}\Psi_{1})&\partial^{-1}_{q}(\Psi_{1})\\
\Phi_{1}&\Phi_{2}&1\\
\partial_{q}(\Phi_{1})&\partial_{q}(\Phi_{2})&0
\end{vmatrix}\neq0,\nonumber
\end{eqnarray}
we have gauge transformations
\begin{eqnarray}
T_{2+1}&=&\frac{1}{IW^{q}_{1,2+1}}
\begin{vmatrix}
\partial^{-1}_{q}(\Phi_{1}\Psi_{1})& \partial^{-1}_{q}(\Phi_{2}\Psi_{1})&\partial^{-1}_{q}\circ\Psi_{1}\\
\Phi_{1}&\Phi_{2}&1\\
\partial_{q}(\Phi_{1})&\partial_{q}(\Phi_{2})&\partial_{q}
\end{vmatrix}\nonumber\\
&=&a_{-1}\circ\partial^{-1}_{q}\circ\Psi_{1}+a_{0}+a_{-1}\circ\partial_{q},
\end{eqnarray}
\begin{eqnarray}
T^{-1}_{2+1}&=&-\frac{q^{-1}IW^{q}_{1,2+1}}{IW^{q}_{1,2}\theta(IW^{q}_{1,2})}
\begin{vmatrix}
\Phi_{1}\partial^{-1}_{q}&\partial^{-1}_{q}
\theta(\Psi_{1}\Phi_{1})\\
\Phi_{2}\partial^{-1}_{q}&\partial^{-1}_{q}
\theta(\Psi_{1}\Phi_{2})
\end{vmatrix}\nonumber\\
&=&\Phi_{1}\partial^{-1}_{q}\circ b_{1}+\Phi_{2}\partial^{-1}_{q}\circ b_{2}.
\end{eqnarray}

Using the Theorem 1, we can get the $q$-cmKP hierarchy generated by $T_{2+1}$
\begin{eqnarray}
(L^{l})_{\le0}&=&(T_{2+1}\circ\partial^{l}_{q}\circ T_{2+1}^{-1})_{\le0}\nonumber\\
&=&-\sum_{j=1}^{2}T_{2+1}(\partial^{l}_{q}(\Phi_{j}))\circ\partial^{-1}_{q}\circ \theta(\partial^{-1}_{q}(b_{j}))\circ\partial_{q}\nonumber\\
&&+(-1)^{l}q^{-(1+2+\cdots+l)}a_{1}\circ\partial^{-1}_{q}\circ[(T_{2+1}^{-1}\partial^{-1}_{q})(\theta^{-l}
(\partial^{l}_{q}(\Psi_{1})))]\circ\partial_{q}\nonumber\\
&&+\sum_{j=1}^{2}a_{1}\partial^{-1}_{q}(\Psi_{1}\partial^{l}_{q}(\Phi_{j}\partial^{-1}_{q}
(b_{j}))),
\nonumber\\
(L^{l})_{\le0}&=&(L_{\alpha})_{\le0}+(L_{\beta})_{\le0}+(L_{\gamma})_{\le0},\nonumber
\end{eqnarray}
here $\alpha=1$, $\beta=1$. We have
\begin{eqnarray}
IW^{q}_{k,n+1+\alpha+1}&=&IW^{q}_{1,5}(\Psi_{1},\Phi_{1},\Phi_{2},\partial^{l}_{q}(\Phi_{1}),\partial^{l}_{q}(\Phi_{2}),1)\nonumber\\
&=&\begin{vmatrix}
\partial^{-1}_{q}(\Phi_{1}\Psi_{1})& \partial^{-1}_{q}(\Phi_{2}\Psi_{1})&\partial^{-1}_{q}(\Psi_{1}\partial^{l}_{q}(\Phi_{1}))&\partial^{-1}_{q}(\Psi_{1}\partial^{l}_{q}(\Phi_{2}))&\partial^{-1}_{q}(\Psi_{1})\\
\Phi_{1}&\Phi_{2}&\partial^{l}_{q}(\Phi_{1})&\partial^{l}_{q}(\Phi_{2})&1\\
\partial_{q}(\Phi_{1})&\partial_{q}(\Phi_{2})&\partial^{l+1}_{q}(\Phi_{1})&\partial^{l+1}_{q}(\Phi_{2})&0\\
\partial^{2}_{q}(\Phi_{1})&\partial^{2}_{q}(\Phi_{2})&\partial^{l+2}_{q}(\Phi_{1})&\partial^{l+2}_{q}(\Phi_{2})&0\\
\partial^{3}_{q}(\Phi_{1})&\partial^{3}_{q}(\Phi_{2})&\partial^{l+3}_{q}(\Phi_{1})&\partial^{l+3}_{q}(\Phi_{2})&0
\end{vmatrix}\nonumber\\
&=&0,\nonumber
\end{eqnarray}
here we let
\begin{eqnarray}
\Phi_{1}=e_{q}(\mu_{1}x)e^{\theta_{1}}+e_{q}(\mu_{2}x)e^{\theta_{2}},\nonumber\\
\Phi_{2}=e_{q}(\mu_{3}x)e^{\theta_{3}}+e_{q}(\mu_{4}x)e^{\theta_{4}},\nonumber\\
\Psi_{1}=e_{\frac{1}{q}}(-\lambda_{1}x)e^{\eta_{1}}+e_{\frac{1}{q}}(-\lambda_{2}x)e^{\eta_{2}},\nonumber
\end{eqnarray}
and
\begin{eqnarray}
&&\theta_{i}=c_{i}+\mu_{i}t_{1}+\mu^{2}_{i}t_{2}+\mu^{3}_{i}t_{3}+\cdots+\mu^{n}_{i}t_{n}+\cdots,\quad\quad i=1,2,3,4,\nonumber\\
&&\eta_{j}=\rho_{j}-\frac{1}{q}\lambda_{j}t_{1}-\frac{1}{q^{2}}\lambda^{2}_{j}t_{2}-\frac{1}{q^{3}}\lambda^{3}_{j}t_{3}-\cdots-\frac{1}{q^{n}}\lambda^{n}_{j}t_{n}-\cdots,\quad\quad j=1,2, \nonumber
\end{eqnarray}
where $c_{i}$, $\rho_{j}$ are constants. We have
\begin{eqnarray}
&&IW^{q}_{1,5}(\Psi_{1},\Phi_{1},\Phi_{2},\partial^{l}_{q}(\Phi_{1}),\partial^{l}_{q}(\Phi_{2}),1)\nonumber\\
&&=\begin{vmatrix}
\partial^{-1}_{q}(\Phi_{1}\Psi_{1})&\partial^{-1}_{q}(\Phi_{2}\Psi_{1})&\partial^{-1}_{q}(\Psi_{1}\partial^{l}_{q}(\Phi_{1}))&\partial^{-1}_{q}(\Psi_{1}\partial^{l}_{q}(\Phi_{2}))&\partial^{-1}_{q}(\Psi_{1})\\
\Phi_{1}&\Phi_{2}&\partial^{l}_{q}(\Phi_{1})&\partial^{l}_{q}(\Phi_{2})&1\\
\partial_{q}(\Phi_{1})&\partial_{q}(\Phi_{2})&\partial^{l+1}_{q}(\Phi_{1})&\partial^{l+1}_{q}(\Phi_{2})&0\\
\partial^{2}_{q}(\Phi_{1})&\partial^{2}_{q}(\Phi_{2})&\partial^{l+2}_{q}(\Phi_{1})&\partial^{l+2}_{q}(\Phi_{2})&0\\
\partial^{3}_{q}(\Phi_{1})&\partial^{3}_{q}(\Phi_{2})&\partial^{l+3}_{q}(\Phi_{1})&\partial^{l+3}_{q}(\Phi_{2})&0
\end{vmatrix}\nonumber\\
&&=-\begin{vmatrix}
\partial^{-1}_{q}(\Phi_{1}\Psi_{1})& \partial^{-1}_{q}(\Phi_{2}\Psi_{1})&\partial^{-1}_{q}(\Psi_{1}\partial^{l}_{q}(\Phi_{1}))&\partial^{-1}_{q}(\Psi_{1}\partial^{l}_{q}(\Phi_{2}))\\
\partial_{q}(\Phi_{1})&\partial_{q}(\Phi_{2})&\partial^{l+1}_{q}(\Phi_{1})&\partial^{l+1}_{q}(\Phi_{2})\\
\partial^{2}_{q}(\Phi_{1})&\partial^{2}_{q}(\Phi_{2})&\partial^{l+2}_{q}(\Phi_{1})&\partial^{l+2}_{q}(\Phi_{2})\\
\partial^{3}_{q}(\Phi_{1})&\partial^{3}_{q}(\Phi_{2})&\partial^{l+3}_{q}(\Phi_{1})&\partial^{l+3}_{q}(\Phi_{2})
\end{vmatrix}\nonumber\\
&&+\partial^{-1}_{q}(\Psi_{1})\begin{vmatrix}
\Phi_{1}&\Phi_{2}&\partial^{l}_{q}(\Phi_{1})&\partial^{l}_{q}(\Phi_{2})\\
\partial_{q}(\Phi_{1})&\partial_{q}(\Phi_{2})&\partial^{l+1}_{q}(\Phi_{1})&\partial^{l+1}_{q}(\Phi_{2})\\
\partial^{2}_{q}(\Phi_{1})&\partial^{2}_{q}(\Phi_{2})&\partial^{l+2}_{q}(\Phi_{1})&\partial^{l+2}_{q}(\Phi_{2})\\
\partial^{3}_{q}(\Phi_{1})&\partial^{3}_{q}(\Phi_{2})&\partial^{l+3}_{q}(\Phi_{1})&\partial^{l+3}_{q}(\Phi_{2})
\end{vmatrix}\nonumber\\
&&=-\partial^{-1}_{q}(\Phi_{1}\Psi_{1})\begin{vmatrix}
\partial_{q}(\Phi_{2})&\partial^{l+1}_{q}(\Phi_{1})&\partial^{l+1}_{q}(\Phi_{2})\\
\partial^{2}_{q}(\Phi_{2})&\partial^{l+2}_{q}(\Phi_{1})&\partial^{l+2}_{q}(\Phi_{2})\\
\partial^{3}_{q}(\Phi_{2})&\partial^{l+3}_{q}(\Phi_{1})&\partial^{l+3}_{q}(\Phi_{2})
\end{vmatrix}\nonumber\\
&&+\partial^{-1}_{q}(\Phi_{2}\Psi_{1})\begin{vmatrix}
\partial_{q}(\Phi_{1})&\partial^{l+1}_{q}(\Phi_{1})&\partial^{l+1}_{q}(\Phi_{2})\\
\partial^{2}_{q}(\Phi_{1})&\partial^{l+2}_{q}(\Phi_{1})&\partial^{l+2}_{q}(\Phi_{2})\\
\partial^{3}_{q}(\Phi_{1})&\partial^{l+3}_{q}(\Phi_{1})&\partial^{l+3}_{q}(\Phi_{2})
\end{vmatrix}\nonumber\\
&&-\partial^{-1}_{q}(\Psi_{1}\partial^{l}_{q}(\Phi_{1}))\begin{vmatrix}
\partial_{q}(\Phi_{1})&\partial_{q}(\Phi_{2})&\partial^{l+1}_{q}(\Phi_{2})\\
\partial^{2}_{q}(\Phi_{1})&\partial^{2}_{q}(\Phi_{2})&\partial^{l+2}_{q}(\Phi_{2})\\
\partial^{3}_{q}(\Phi_{1})&\partial^{3}_{q}(\Phi_{2})&\partial^{l+3}_{q}(\Phi_{2})
\end{vmatrix}\nonumber\\
&&+\partial^{-1}_{q}(\Psi_{1}\partial^{l}_{q}(\Phi_{2}))\begin{vmatrix}
\partial_{q}(\Phi_{1})&\partial_{q}(\Phi_{2})&\partial^{l+1}_{q}(\Phi_{1})\\
\partial^{2}_{q}(\Phi_{1})&\partial^{2}_{q}(\Phi_{2})&\partial^{l+2}_{q}(\Phi_{1})\\
\partial^{3}_{q}(\Phi_{1})&\partial^{3}_{q}(\Phi_{2})&\partial^{l+3}_{q}(\Phi_{1})
\end{vmatrix}\nonumber\\
&&+\partial^{-1}_{q}(\Psi_{1})\begin{vmatrix}
\Phi_{1}&\Phi_{2}&\partial^{l}_{q}(\Phi_{1})&\partial^{l}_{q}(\Phi_{2})\\
\partial_{q}(\Phi_{1})&\partial_{q}(\Phi_{2})&\partial^{l+1}_{q}(\Phi_{1})&\partial^{l+1}_{q}(\Phi_{2})\\
\partial^{2}_{q}(\Phi_{1})&\partial^{2}_{q}(\Phi_{2})&\partial^{l+2}_{q}(\Phi_{1})&\partial^{l+2}_{q}(\Phi_{2})\\
\partial^{3}_{q}(\Phi_{1})&\partial^{3}_{q}(\Phi_{2})&\partial^{l+3}_{q}(\Phi_{1})&\partial^{l+3}_{q}(\Phi_{2})
\end{vmatrix},\nonumber
\end{eqnarray}
and
\begin{eqnarray}
&&\begin{vmatrix}
\partial_{q}(\Phi_{2})&\partial^{l+1}_{q}(\Phi_{1})&\partial^{l+1}_{q}(\Phi_{2})\\
\partial^{2}_{q}(\Phi_{2})&\partial^{l+2}_{q}(\Phi_{1})&\partial^{l+2}_{q}(\Phi_{2})\\
\partial^{3}_{q}(\Phi_{2})&\partial^{l+3}_{q}(\Phi_{1})&\partial^{l+3}_{q}(\Phi_{2})
\end{vmatrix}\nonumber\\
&&=\mu^{l+1}_{1}(\mu^{l}_{3}-\mu^{l}_{4})
\mu_{3}\mu_{4}e_{q}(\mu_{1}x)e_{q}(\mu_{3}x)e_{q}(\mu_{4}x)e^{\theta_{1}+\theta_{3}+\theta_{4}}
\begin{vmatrix}
1&1&1\\
\mu_{1}&\mu_{3}&\mu_{4}\\
\mu_{1}^{2}&\mu_{3}^{2}&\mu_{4}^{2}
\end{vmatrix}\nonumber\\
&&+\mu^{l+1}_{2}(\mu^{l}_{3}-\mu^{l}_{4})
\mu_{3}\mu_{4}e_{q}(\mu_{2}x)e_{q}(\mu_{3}x)e_{q}(\mu_{4}x)e^{\theta_{2}+\theta_{3}+\theta_{4}}
\begin{vmatrix}
1&1&1\\
\mu_{2}&\mu_{3}&\mu_{4}\\
\mu_{2}^{2}&\mu_{3}^{2}&\mu_{4}^{2}
\end{vmatrix},\nonumber
\end{eqnarray}
when $\mu_{3}=\mu_{4},c_{3}=c_{4}$, we know $\theta_{3}=\theta_{4}$, thus
\begin{eqnarray}
&&\begin{vmatrix}
\partial_{q}(\Phi_{2})&\partial^{l+1}_{q}(\Phi_{1})&\partial^{l+1}_{q}(\Phi_{2})\\
\partial^{2}_{q}(\Phi_{2})&\partial^{l+2}_{q}(\Phi_{1})&\partial^{l+2}_{q}(\Phi_{2})\\
\partial^{3}_{q}(\Phi_{2})&\partial^{l+3}_{q}(\Phi_{1})&\partial^{l+3}_{q}(\Phi_{2})
\end{vmatrix}=0.\nonumber
\end{eqnarray}

Similarly, when $\mu_{1}=\mu_{2},c_{1}=c_{2}$, we know $\theta_{1}=\theta_{2}$, thus
\begin{eqnarray}
\begin{vmatrix}
\partial_{q}(\Phi_{1})&\partial^{l+1}_{q}(\Phi_{1})&\partial^{l+1}_{q}(\Phi_{2})\\
\partial^{2}_{q}(\Phi_{1})&\partial^{l+2}_{q}(\Phi_{1})&\partial^{l+2}_{q}(\Phi_{2})\\
\partial^{3}_{q}(\Phi_{1})&\partial^{l+3}_{q}(\Phi_{1})&\partial^{l+3}_{q}(\Phi_{2})
\end{vmatrix}=0,\nonumber
\end{eqnarray}
and
\begin{eqnarray}
\partial^{-1}_{q}(\Psi_{1}\partial^{l}_{q}(\Phi_{1}))\begin{vmatrix}
\partial_{q}(\Phi_{1})&\partial_{q}(\Phi_{2})&\partial^{l+1}_{q}(\Phi_{2})\\
\partial^{2}_{q}(\Phi_{1})&\partial^{2}_{q}(\Phi_{2})&\partial^{l+2}_{q}(\Phi_{2})\\
\partial^{3}_{q}(\Phi_{1})&\partial^{3}_{q}(\Phi_{2})&\partial^{l+3}_{q}(\Phi_{2})
\end{vmatrix}=0,\begin{vmatrix}
\partial_{q}(\Phi_{1})&\partial_{q}(\Phi_{2})&\partial^{l+1}_{q}(\Phi_{1})\\
\partial^{2}_{q}(\Phi_{1})&\partial^{2}_{q}(\Phi_{2})&\partial^{l+2}_{q}(\Phi_{1})\\
\partial^{3}_{q}(\Phi_{1})&\partial^{3}_{q}(\Phi_{2})&\partial^{l+3}_{q}(\Phi_{1})
\end{vmatrix}=0.\nonumber
\end{eqnarray}
From the above equations, we have let $\theta_{1}=\theta_{2},\theta_{3}=\theta_{4}$, so we can get that
\begin{eqnarray}
\tau^{q}_{2+1}&=&IW^{q}_{1,2}\nonumber\\
&=&\begin{vmatrix}
\partial^{-1}_{q}(\Phi_{1}\Psi_{1})& \partial^{-1}_{q}(\Phi_{2}\Psi_{1})\\
\Phi_{1}&\Phi_{2}
\end{vmatrix}\nonumber\\
&=&2\partial^{-1}_{q}(\Phi_{1}\Psi_{1})e_{q}(\mu_{3}x)e^{\theta_{3}}-2\partial^{-1}_{q}(\Phi_{2}\Psi_{1})e_{q}(\mu_{1}x)e^{\theta_{1}}.\nonumber
\end{eqnarray}
 After a series of calculations, the following expressions $T_{2+1}(\partial^{l}_{q}(\Phi_{1})), T_{2+1}(\partial^{l}_{q}(\Phi_{2})),b_{1},b_{2}$,\\
 $a_{-1},(T_{n+k}^{-1}\partial^{-1}_{q})
 ^{*}(\theta^{-l}
 (\partial^{l}_{q}(\Psi_{1})))$
 are obtained,
 \begin{eqnarray}
 T_{2+1}(\partial^{l}_{q}(\Phi_{1}))
 =\frac{4}{IW^{q}_{1,3}}(-\mu^{l}_{1}\partial^{-1}_{q}(\Phi_{1}\Psi_{1})+\partial^{-1}_{q}(\Psi_{1}\partial^{l}_{q}(\Phi_{1})))e_{q}(\mu_{1}x)e_{q}(\mu_{3}x)e^{\theta_{1}+\theta_{3}}
 \begin{vmatrix}
 1&1\\\mu_{1}&\mu_{3}
 \end{vmatrix},\nonumber
 \end{eqnarray}
 \begin{eqnarray}
 T_{2+1}(\partial^{l}_{q}(\Phi_{2}))
 =\frac{4}{IW^{q}_{1,3}}(-\mu^{l}_{3}\partial^{-1}_{q}(\Phi_{2}\Psi_{1})+\partial^{-1}_{q}(\Psi_{1}\partial^{l}_{q}(\Phi_{2})))e_{q}(\mu_{1}x)e_{q}(\mu_{3}x)e^{\theta_{1}+\theta_{3}}
 \begin{vmatrix}
 1&1\\\mu_{1}&\mu_{3}
 \end{vmatrix},\nonumber
 \end{eqnarray}
 \begin{eqnarray}
b_{1}&=&-\frac{q^{-1}IW^{q}_{1,3}(\Psi_{1},\Phi_{1},\Phi_{2},1)\partial^{-1}_{q}(\theta(\Psi_{1}\Phi_{2}))}{IW^{q}_{1,2}(
\Psi_{1},\Phi_{1},\Phi_{2})\theta(IW^{q}_{1,2}
	(\Psi_{1},\Phi_{1},\Phi_{2}))},\nonumber\\
b_{2}&=&\frac{q^{-1}IW^{q}_{1,3}(\Psi_{1},\Phi_{1},\Phi_{2},1)\partial^{-1}_{q}(\theta(\Psi_{1}\Phi_{1}))}{IW^{q}_{1,2}(
	\Psi_{1},\Phi_{1},\Phi_{2})\theta(IW^{q}_{1,2}
	(\Psi_{1},\Phi_{1},\Phi_{2}))},\nonumber
 \end{eqnarray}
 \begin{eqnarray}
 a_{-1}&=&\frac{1}{IW^{q}_{1,3}}
 \begin{vmatrix}\Phi_{1}&\Phi_{2}\\
 \partial_{q}(\Phi_{1})&\partial_{q}(\Phi_{2})
 \end{vmatrix}\nonumber\\
 &=&\frac{4}{IW^{q}_{1,3}}(\mu_{3}-\mu_{1})e_{q}(\mu_{1}x)e_{q}(\mu_{3}x)e^{\theta_{1}+\theta_{3}}.\nonumber
 \end{eqnarray}

%\begin{eqnarray}
%(T_{n+k}^{-1})^{*}(\theta^{-l}
%(\partial^{l}_{q}(\Psi_{1})))=\frac{q^{-1}IW^{q}_{1,2+1}}{IW^{q}_{1,2}\theta(IW^{q}_{1,2})}
%\begin{vmatrix}
%\theta\partial^{-1}_{q}(\Phi_{1}\theta^{-l}
%(\partial^{l}_{q}(\Psi_{1})))&\partial^{-1}_{q}
%\theta(\Psi_{1}\Phi_{1})\\
%\theta\partial^{-1}_{q}(\Phi_{2}\theta^{-l}
%(\partial^{l}_{q}(\Psi_{1})))&\partial^{-1}_{q}
%\theta(\Psi_{1}\Phi_{2})
%\end{vmatrix}
%\end{eqnarray}
%so
%\begin{eqnarray}
%(T_{n+k}^{-1}\partial^{-1}_{q})
%^{*}
%(\theta^{-l}
%(\partial^{l}_{q}(\Psi_{1})))=-\theta\partial^{-1}_{q}(\frac{q^{-1}IW^{q}_{1,2+1}}{IW^{q}_{1,2}\theta(IW^{q}_{1,2})}
%\begin{vmatrix}
%\theta\partial^{-1}_{q}(\Phi_{1}\theta^{-l}
%(\partial^{l}_{q}(\Psi_{1})))&\partial^{-1}_{q}
%\theta(\Psi_{1}\Phi_{1})\\
%\theta\partial^{-1}_{q}(\Phi_{2}\theta^{-l}
%(\partial^{l}_{q}(\Psi_{1})))&\partial^{-1}_{q}
%\theta(\Psi_{1}\Phi_{2})
%\end{vmatrix})
%\end{eqnarray}
The goal is to construct $(L^{l})_{\le0}=s_{1}\circ\partial^{-1}_{q}\circ r_{1}\circ\partial_{q}+s_{2}\circ\partial^{-1}_{q}\circ r_{2}\circ\partial_{q}+s_{3}\circ\partial^{-1}_{q}\circ r_{3}\circ\partial_{q}$.
Due to \eqref{L24},
\begin{eqnarray}
r_{3}&=&1,\nonumber\\
s_{3}&=&\sum_{j=1}^{2}a_{1}\partial^{-1}_{q}(\Psi_{1}\partial^{l}_{q}(\Phi_{j}\partial^{-1}_{q}
(b_{j}))).\nonumber
\end{eqnarray}
The formula \eqref{sss} reads as
\begin{eqnarray}
T_{2+1}(\partial^{l}_{q}(\Phi_{1}))=-c_{11}s_{1},\nonumber\\
T_{2+1}(\partial^{l}_{q}(\Phi_{2}))=-c_{21}s_{1}.\nonumber
\end{eqnarray}
 Then we have
\begin{eqnarray}
s_{1}=-\frac{1}{c_{11}}T_{2+1}(\partial^{l}_{q}(\Phi_{1}))=-\frac{1}{c_{21}}T_{2+1}(\partial^{l}_{q}(\Phi_{1})),\nonumber
\end{eqnarray}
here
\begin{eqnarray}
c_{11}=\frac{1}{\mu^{l}_{3}\partial^{-1}_{q}(\Phi_{2}\Psi_{1})-\partial^{-1}_{q}(\Psi_{1}\partial^{l}_{q}(\Phi_{2}))},\nonumber\\
c_{21}=\frac{1}{\mu^{l}_{1}\partial^{-1}_{q}(\Phi_{1}\Psi_{1})-\partial^{-1}_{q}(\Psi_{1}\partial^{l}_{q}(\Phi_{1}))}.\nonumber
\end{eqnarray}
We can get
\begin{eqnarray}
r_{1}=c_{11}\theta(\partial^{-1}_{q}(b_{1}))+c_{21}\theta(\partial^{-1}_{q}(b_{2}))\nonumber
\end{eqnarray}
by the last equation of the formula \eqref{sss}.
$r_{2}$ and $s_{2}$ are obvious, because before the group reduction, it only has one component, we have
\begin{eqnarray}
s_{2}&=&(-1)^{l}q^{-(1+2+\cdots+l)}a_{-1}\nonumber\\
&=&(-1)^{l}q^{-(1+2+\cdots+l)}\frac{4}{IW^{q}_{1,3}}(\mu_{3}-\mu_{1})e_{q}(\mu_{1}x)e_{q}(\mu_{3}x)e^{\theta_{1}+\theta_{3}},\nonumber\\
r_{2}&=&(T_{n+k}^{-1}\partial^{-1}_{q})
^{*}
(\theta^{-l}
(\partial^{l}_{q}(\Psi_{1})))\nonumber\\
&=&-\theta\partial^{-1}_{q}\left(\frac{q^{-1}IW^{q}_{1,2+1}}{IW^{q}_{1,2}\theta(IW^{q}_{1,2})}
\begin{vmatrix}
\theta\partial^{-1}_{q}(\Phi_{1}\theta^{-l}
(\partial^{l}_{q}(\Psi_{1})))&\partial^{-1}_{q}
\theta(\Psi_{1}\Phi_{1})\\
\theta\partial^{-1}_{q}(\Phi_{2}\theta^{-l}
(\partial^{l}_{q}(\Psi_{1})))&\partial^{-1}_{q}
\theta(\Psi_{1}\Phi_{2})
\end{vmatrix}\right).\nonumber
\end{eqnarray}

\section{CONCLUSION}
In this paper, the $l$-constrained $q$-mKP hierarchy generated by the gauge transformation operator $T_{n+k}(n\geq k)$ is divided into three parts. Therefore, we divide the Lax operator of the $M$-component $q$-cmKP hierarchy into three groups by grouping reduction
\begin{eqnarray}
(L^{l})_{\leq0}=\sum_{i=1}
^{\alpha}s_{i}\partial^{-1}_{q}r_{i}\partial_{q}+\sum_{i=\alpha+1}
^{\alpha+\beta=M-1}s_{i}\partial^{-1}_{q}r_{i}\partial_{q}+s_{M}\partial^{-1}_{q}r_{M}\partial_{q}.\nonumber
 \end{eqnarray}
 In Proposition 8, we give the necessary and sufficient condition for reducing the generalised $q$-Wronskian solutions of the $q$-mKP hierarchy to the $q$-cmKP hierarchy. Finally, we give a concrete example by letting $T_{n+1}$ and $M=3$.

We give the form of the $l$-constrained mKP hierarchy generated by gauge transformation operator $T_{n+k}$,
\begin{eqnarray}
(L^{l})_{\le0}&=&-\sum_{j=1}^{n}T_{n+k}(\Phi_{j}^{(l)})\circ\partial^{-1}\circ \int b_{j}dx\circ\partial+(-1)^{l}\sum_{p=-1}^{-k}a_{p}\circ\partial^{-1}\circ[(T_{n+k}^{-1}\partial^{-1})^{*}(\Psi_{|p|}^{(l)})]\circ\partial\nonumber\\
&&+\sum_{j=1}^{n}\sum_{p=-1}^{-k}a_{p}\int(\Phi_{j}\int b_{j}dx)^{(l)}\Psi_{|p|}dx.\nonumber
\end{eqnarray}
 When $q\to1$, the results can be returned to the classical cmKP hierarchy. This further verifies the correctness of this paper.
\bigskip
\bigskip

\textbf{Acknowledgements:} This work is supported by the National Natural Science Foundation of China under Grant Nos.12171133, 12171132 and 11871446, and the Anhui Province Natural Science Foundation (No. 2008085MA05).

\end{document}